\documentclass[pra,groupaddress,twocolumn]{revtex4}
\usepackage{graphicx}
\usepackage{amsmath}
\usepackage{bm}
\begin{document} \title{Hydrodynamic Tensor-DFT with correct susceptibility}
\author{Igor V. Ovchinnikov}
\author{Lizette A. Bartell}
\author{Daniel Neuhauser}\email{dxn@chem.ucla.edu}
\affiliation{Department of Chemistry and Biochemistry, University of
California at Los Angeles, Los Angeles, CA, 90095-1569}
\begin{abstract}
In a previous work we developed a family of orbital-free tensor
equations for DFT [J. Chem. Phys. 124, 024105 (2006)].  The theory
is a combination of the coupled hydrodynamic moment equations
hierarchy with a cumulant truncation of the one-body electron
density matrix.  A basic ingredient in the theory is how to truncate
the series of equation of motion for the moments.  In the original
work we assumed that the cumulants vanish above a certain order (N).
Here we show how to modify this assumption to obtain the correct
susceptibilities.  This is done for N=3, a level above the previous
study.  At the desired truncation level a few relevant terms are
added, which, with the right combination of coefficients, lead to
excellent agreement with the Kohn-Sham Lindhard susceptibilities for
an uninteracting system.  The approach is also powerful away from
linear response, as demonstrated in a non-perturbative study of a
jellium with a repulsive core, where excellent matching with
Kohn-Sham simulations is obtained while the Thomas Fermi and
von-Weiszacker methods show significant deviations. In addition,
time-dependent linear response studies at the new N=3 level
demonstrate our previous assertion that as the order of the theory
is increased, new additional transverse sound modes appear mimicking
the RPA transverse dispersion region.
\end{abstract}
\maketitle

\section{Introduction}
The development of new methods for quantum dynamics based upon
hydrodynamic representations is very promising. In hydrodynamics the
kinetics of the system is defined by a lesser number of variables
than the number of variables required to define the complete
one-particle density matrix (which contains all the information on
off-diagonal quantum coherence as in, \emph{e.g.}, the Kohn-Sham
approach). For stationary studies the hydrodynamics approach is
related to orbital-free density-functional theory
\cite{books,Carter1,Carter2,Carter3,GarciaGonzalez,Smargiassi,
Wang,Watson,Nehete,Aguado,Govind,Depristo,Wang1,Brack,Frankcombe,Sim,Ayers,Chan,Choly,Glossman,Gross,Lee,Plindov,Yang,Trickey}.
It is the reduced number of variables depicting the system that
makes hydrodynamical theories applicable for numerical studies of
relatively large systems.

The simplest hydrodynamical approach is similar to the de
Broglie-Bohm formulation of one-particle quantum mechanics
\cite{generalMethods9,Derrickson,Madelung,Deb,Lopreore,Wyatt,Burant,DayMultyDim}.
In this approximation the complete complex-valued one-particle
density matrix is substituted by two real valued fields $\rho$ and
$\phi$, which are combined in an order parameter
$\psi=\sqrt{\rho}\exp(i\varphi)$. The equations of motion are
obtained by minimizing a Ginzburg-Landau-like functional on $\psi$.
In addition the density matrix is assumed to possess long-range
off-diagonal one-particle correlations.

A more rigorous and asymptotically exact approach is an infinite
hierarchy of coupled hydrodynamic moment (CHM) equations
\cite{Moments0,Moments2,Moments3,Moments4,Moments5}. The moments
come from a Taylor expansion of the one-particle density matrix with
respect to the off-diagonal variable. To get a tractable system of
equations the infinite hierarchy must be truncated. The most
physically meaningful truncation is a cumulant expansion for the
density matrix \cite{Moments0}. Specifically, one decides on an
order to terminate the method at; a low order will be less
numerically demanding but less accurate than a higher one. Then, at
that order, labeled N, the (N+1)-th order moment is expanded in
terms of the previous set of moments, through the use of the
cumulant expansion.

The CHM theory and the accompanying cumulant truncation have been
applied so far to systems where particle statistics does not play an
important role.  In Ref.\cite{ours} we have generalized the CHM
theory and cumulant expansion to statistically degenerate fermions.
The main point has been the modification of the unperturbed
one-particle density matrix of a locally homogenous electron gas by
using the cumulants. Since the approach uses successive tensors, we
labeled it Hydrodynamic tensor DFT (HTDFT).

It turns out that the lowest level of truncation, $N = 1$, HTDFT
corresponds to a de-Brogilie-Bohm quantum hydrodynamics and in
addition naturally incorporates the Thomas-Fermi \cite{Thomas}
kinetic energy term into the energy functional. At the next level,
$N=2$, HTDFT starts reproducing the spectrum of a homogenous Fermi
liquid, \emph{i.e.}, it gets transverse excitations, rather than
just classical plasmonic longitudinal excitations.  The transverse
sound mode mimics the elementary excitations' density of states.

A crucial feature of HTDFT is the value of the cumulant used at the
truncation. In Ref.\cite{ours} we assumed that the $(N+1)$-th order
cumulant is zero.  It turns out, however, as we show here, that this
assumption leads to a wrong susceptibility for a homogenous electron
gas, \emph{i.e.}, to a wrong linear response to a perturbation, even
for a non-interacting system of electrons.  We show here how to
remedy this problem.  This is exemplified below for truncation at
the $N = 3$ level, which is the first level where the method will
yield different ground-state results from the Thomas-Fermi approach.
Specifically, the 4'th order cumulant is written as a sum of terms
involving the gradients of the previous moments.  The coefficients
of these terms are obtained by fitting to the exact susceptibility
of a non-interacting set of electrons (the Lindhard function).

The balance of the paper is as follows. The general methodology is
first developed in Section II. In Section III the derivation of a
correct susceptibility is done. Section IV applies the methodology
to a static non-perturbative numerical study of a jellium with a
deep spherically symmetric hole, where we show that the agreement
with Kohn-Sham results is excellent while the Thomas Fermi and the
von-Weiszacker methods have significant errors. Section V is a
linear-response time-dependent study of the approach for N=3 as a
function of frequency and wavevector. This latter part is a direct
continuation of our work in Ref.\cite{ours} for $N=2$, and proves
that there is an additional sound mode with respect to the $N=2$
case, just as suggested in Ref.\cite{ours}. Conclusions follow in
Section VI.

\section{The system and Tensor-DFT formulation}

\subsection{Coupled Hydrodynamic Moment Hierarchy}

For completeness, we rederive the basic aspects of the theory (see
Ref.\cite{ours}). We assume that the many electron system can be
described by the one-particle density matrix, $\rho^{(1)}$.  The
one-electron Hamiltonian governing this system, $h$, is, as usual,
composed of kinetic terms, and a local potential terms.

The one-particle density matrix is then expressed in terms of
average and difference coordinates as:
\begin{eqnarray}
\rho^{(1)}(\bm R, \bm s) = \langle \hat \psi^\dagger(\bm R - \bm
s/2) \hat \psi(\bm R + \bm s/2)\rangle.
\end{eqnarray}
The time evolution of the one-particle density matrix is governed by
the Heisenberg equation, $i\dot\rho=\left[h,\rho\right]$, which in
those coordinates takes the form:
\begin{eqnarray}
i\frac{\partial }{\partial t} \rho^{(1)}(\bm R, \bm s) &=& \hat
P_\alpha \hat p_\alpha
\rho^{(1)}\nonumber \\
&&+\left(\tilde V(\bm R+\bm s/2)-\tilde V(\bm R - \bm
s/2)\right)\rho^{(1)}.\label{basicequation}
\end{eqnarray}
Here $\hat P_\alpha$ and $\hat p_\alpha$ stand for the derivatives
over the coordinates $R_\alpha$ and $s_\alpha$
\begin{eqnarray}
\hat P_\alpha = -i \partial/\partial R_\alpha,&& \hat p_\alpha = -i \partial/\partial s_\alpha,
\end{eqnarray}
and $\tilde V(R)$ is the effective potential, which also takes into account the two-body interactions:
\begin{eqnarray}
\label{EffectivePotential} \tilde V(\bm R) = \int \frac{\rho(\bm
R')-\rho_0(\bm R')}{|\bm R - \bm R'|}d^3R' - \frac{\delta
E_{xc}}{\delta \rho(\bm R)} + V_{ext}(\bm R),
\end{eqnarray}
where $\rho(\bm R)=\rho^{(1)}(\bm R,\bm 0)$ is the spatial electron
density, $\rho_0(\bm R)$ is the positive nuclear charge density,
$V_{ext}$ is any external potential, and $E_{xc}$ is the
exchange-correlation energy and $\rho_0$ is the nuclei density.
There are a variety of functions $V_{xc}\equiv\delta
E_{xc}/\delta\rho$ in the literature (see, \emph{e.g.},
Ref.\cite{Carter1,Carter2,Carter3}). For us, however, the specific
form of $V_{xc}$ is not important. [In future works we will aim to
derive a form of $V_{xc}$ which depends also on other moments in
addition to $\rho(\bm R)$.]

The particle kinetics in the system can be exactly described by the
complete infinite set of hydrodynamic moments (dynamic tensors)
\cite{Moments0,Moments2,Moments3,Moments4,Moments5,ours}, which are
the derivatives of the one-particle density matrix with respect to
the off-diagonal distance, $\bm s$, at $\bm s = \bm 0$:
\begin{eqnarray}
\Phi^{(N)}_{l_1\dots l_N}(\bm R) = \hat p_{l_1}\dots\hat
p_{l_N}\left.\rho^{(1)}(\bm R, \bm s)\right|_{\bm s=\bm
0}.\label{Phiintroduction}
\end{eqnarray}
The particle and the current spatial densities are merely the first two tensors in the family:
\begin{eqnarray} \Phi^{(0)}(\bm R) = \rho(\bm R)&,& \Phi^{(1)}_{i}(\bm R) = J_i(\bm R).
\end{eqnarray}

By using Eq.(\ref{basicequation}) one derives an infinite set of equations which connects the moments at different orders:
\begin{subequations}
\label{setofequations}
\begin{eqnarray}
\frac\partial{\partial t} \rho &=& -\nabla_\alpha J_\alpha,\\
\frac\partial{\partial t} J_i &=& -\nabla_\alpha \Phi^{(2)}_{i\alpha} -
\rho \nabla_i \tilde V,\\
\frac\partial{\partial t} \Phi^{(2)}_{ik}
&=& -\nabla_\alpha \Phi^{(3)}_{ik\alpha} - J_i \nabla_{k}\tilde V -J_k\nabla_i\tilde V,\\
\frac\partial{\partial t} \Phi^{(3)}_{ikl}
&=& -\nabla_\alpha \Phi^{(4)}_{ikl\alpha} - \Phi^{(2)}_{ik} \nabla_l\tilde V -
\Phi^{(2)}_{il} \nabla_k \tilde V \nonumber
\\&& - \Phi^{(2)}_{lk}\nabla_i \tilde V + \frac14\rho \nabla_i\nabla_k\nabla_l \tilde V,\\
\text{etc.}\nonumber
\end{eqnarray}
\end{subequations}
This generic set of equations is correct for both fermions and
bosons. For this set to be useful one should terminate it at some
level. As usual, this termination is actually a method for
factorizing a moment $\Phi^{(N+1)}$ at some $N$ into moments
$\Phi^{k},k\le N$. In addition, this truncation reflects the Fermi
statistics of the particles. The order $N$ at which one terminates
controls the precision with which we treat the system.

\subsection{Fermi-factorization of higher order dynamic tensors}
\label{FermiFactorization} In Ref.\cite{ours} we proposed a
factorization procedure for the lowest order dynamic tensors
($N=2,3$). Here we describe in detail how the factorization of the
higher order dynamic tensors works in the Fermi case. The method
proposed is based on the following general parametrization of the
one-particle density matrix:
\begin{subequations}
\begin{eqnarray}
\rho^{(1)}(\bm R, \bm s) &=& \rho\exp\{\phi(\bm R,\bm s)\}f_0(\rho,\bm s),\\
\phi(\bm R, \bm s) &=& \sum\limits_{\alpha\ge1}
\frac1{\alpha!}\phi^{(\alpha)}_{i_1i_2\dots i_\alpha}(\bm R)(i
s_{i_1})(i
s_{i_2})\dots (i s_{i_\alpha}),\\ f_0(\rho,\bm s) &=& 3\frac{\sin(k_Fs)-(k_Fs)\cos(k_F s)}{(k_F s)^3},\\
k_F&=&(3\pi^2\rho)^{1/3}.
\end{eqnarray}
\end{subequations}
Here, $f_0$ is the normalized one-particle density matrix of a free
fermion liquid with density $\rho(\bm R)$ and $k_F$ is the local
(density-dependent) Fermi wave-vector. All the cumulants,
$\phi^{(\alpha)}$, are symmetric in all the indices because they are
convolved with the symmetric tensors $s_{i_1}\dots s_{i_n}$ and all
the $\phi^{(\alpha)}$ are real as the one-particle density matrix is
hermitian. The same is true for the tensors $\Phi^{(\alpha)}$.

The physical meaning of this paramerization is as following. If $\phi\equiv 0$, then we end up with the Thomas-Fermi
approximation of a locally homogeneous Fermi liquid. The $\phi$ function perturbs this steady liquid picture, and the tensors
$\phi^{\alpha}$'s and/or $\Phi^{(\alpha)}$'s determine different dynamic characteristics of the flowing electron liquid. The
function $f_0$ assures the Fermi statistics of the particles at the one-particle level.

For brevity we introduce below the tensors
\begin{eqnarray}
{\cal F}^{(\alpha)}\equiv \Phi^{(\alpha)}/\rho,
\end{eqnarray}
instead of $\Phi^{(\alpha)}$. The tensors ${\cal F}^{(\alpha)}$ and
$\phi^{(\alpha)}$ are interrelated. The relations between ${\cal
F}^{(\alpha)}$'s and $\phi^{(\alpha)}$'s for the lowest order
tensors are given below for the first four relations:
\begin{widetext}
\begin{subequations}
\label{relations}
\begin{eqnarray}
{\cal F}^{(1)}&=&\phi^{(1)},\\
{\cal F}^{(2)}&=&\phi^{(2)}+ \overline{\phi^{(1)}\phi^{(1)}} + e^{(2)},\\
{\cal F}^{(3)}&=&\phi^{(3)}+ 3 \overline{\phi^{(1)} \phi^{(2)}}+3 \overline{\phi^{(1)}e^{(2)}}
+\overline{\phi^{(1)}\phi^{(1)}\phi^{(1)}},\\
{\cal F}^{(4)}&=&\phi^{(4)}+ 4 \overline{\phi^{(1)}\phi^{(3)}} +
3{\phi^{(2)}\phi^{(2)}} + 6
\overline{\phi^{(1)}\phi^{(1)}\phi^{(2)}} \nonumber \\&& + 6
\overline{\phi^{(2)} e^{(2)}} + 6
\overline{\phi^{(1)}\phi^{(1)}e^{(2)}}+
\overline{\phi^{(1)}\phi^{(1)}\phi^{(1)}\phi^{(1)}} + e^{(4)},\quad
\label{relations4}
\end{eqnarray}
\end{subequations}
\emph{etc.} Here all terms are absolutely symmetric tensors of their
indices so that there is no need to write down the indices
explicitly; a bar denotes here a complete symmetrization,
\emph{e.g.}, for a product of $a_{i_1\dots i_K}$ and $b_{i_1\dots
i_M}$:
\begin{eqnarray*}
\overline{a b}_{i_1\dots i_{K+M}} =
\frac1{(K+M)!}\sum\limits_{p}a_{p(i_1)\dots
p(i_K)}b_{p(i_{K+1})\dots p(i_{K+M})},
\end{eqnarray*}
where summation is assumed over all $(K+M)!$ permutations of the
indices and $p$ denotes a permutation. The symmetrized
multiplication is associative and it can be considered a
multiplication on a ring of symmetric tensors (note that in
Ref.\cite{ours} a somewhat different symmetrization was used). Here
is an explicit example of the symmetrized multiplication:
\begin{eqnarray}
\label{symmetrization}
\overline{\phi^{(1)}\phi^{(1)}\phi^{(2)}}&\equiv&
\frac16\left(\phi^{(1)}_{i}\phi^{(1)}_{j}\phi^{(2)}_{kl}+
\phi^{(1)}_{i}\phi^{(1)}_{k}\phi^{(2)}_{jl}+\phi^{(1)}_{i}\phi^{(1)}_{l}\phi^{(2)}_{jk}\right.\nonumber\\&&\left.
+\phi^{(1)}_{k}\phi^{(1)}_{l}\phi^{(2)}_{ij}
+\phi^{(1)}_{j}\phi^{(1)}_{l}\phi^{(2)}_{ik}+\phi^{(1)}_{j}\phi^{(1)}_{k}\phi^{(2)}_{il}\right).
\end{eqnarray}
In Eqs. (\ref{relations}) the tensors $e^{(2)}$ and $e^{(4)}$ come
from differentiating the function $f_0$, so that they have the
physical meaning of averaging the particle momenta products over the
unperturbed Fermi sea (ufs):
\begin{eqnarray}
e^{(2)}_{ij} &=& \rho^{-1} \left\langle p_i p_j
\right\rangle_{ufs}=c_{2}\delta_{ij},\\
e^{(4)}_{ijkl}&=&\rho^{-1} \left\langle p_i p_jp_k p_l
\right\rangle_{ufs}=3 c_{4}\overline{\delta\delta}_{ijkl}\equiv
c_{4}\left(\delta_{ij}\delta_{kl}+\delta_{ik}\delta_{jl}+\delta_{il}\delta_{jk}\right),\label{etensors}
\end{eqnarray}
where the kinetic coefficients are defined as
\begin{eqnarray}
c_{2}&=&\frac15 k_F^2, \quad
c_{4}=\frac1{35}k_F^4.\label{definitionofc}
\end{eqnarray}
All the odd order $e$'s vanish.

The general recipe for how to express ${\cal F}^{(N)}$ in terms of $\phi^{(\alpha)}$'s is as follows. ${\cal F}^{(N)}$ is the sum
of all the different symmetrized (in the sense discussed above) products of $\phi^{\alpha}\text{ and } e^{(\alpha)},\alpha\le N$.
An additional rule is that each term may include one (and only one) $e$-tensor.

The relations inverse to Eqs. (\ref{relations}) are:
\begin{subequations}
\label{invrelations}
\begin{eqnarray}
{\cal F}^{(1)}-\phi^{(1)}&=&0\label{invrelations1},\\
{\cal F}^{(2)}-\phi^{(2)}&=&\overline{{\cal F}^{(1)}{\cal F}^{(1)}} + e^{(2)}\label{invrelations2},\\
{\cal F}^{(3)}-\phi^{(3)}&=&3\overline{{\cal F}^{(1)}{\cal F}^{(2)}}
-2\overline{{\cal F}^{(1)}{\cal F}^{(1)}{\cal F}^{(1)}}\label{invrelations3},\\
{\cal F}^{(4)}-\phi^{(4)}&=&4\overline{{\cal F}^{(1)}{\cal
F}^{(3)}}-12\overline{{\cal F}^{(1)}{\cal F}^{(1)}{\cal
F}^{(2)}}+6\overline{{\cal F}^{(1)}{\cal F}^{(1)}{\cal F}^{(1)}{\cal
F}^{(1)}}\nonumber \\ && + 3\overline{{\cal F}^{(2)}{\cal
F}^{(2)}}-3\overline{e^{(2)} e^{(2)}}+e^{(4)}\label{invrelations4}.
\end{eqnarray}
\end{subequations}
\end{widetext}
The inversion of the infinite set of relations (\ref{relations}) is
possible since an expression for any ${\cal F}^{(N)}$ in terms of
$\phi^{(\alpha)}$'s contains only $\phi^{(\alpha)}, \alpha\le N$.
This means that if one knows the expressions for the first $N\quad$
$\phi^{(\alpha)}$ tensors (\emph{e.g.},
Eqs.(\ref{invrelations1}-\ref{invrelations3})) for $N=3$) then by
substituting all the lower order $\phi^{(\alpha)}$'s, $\alpha\le N$,
in the relation for ${\cal F}^{(N+1)}$ (Eq.(\ref{relations4})) with
corresponding expressions in terms of ${\cal F}^{(\alpha)}$'s one
gets the inverse relation for $\phi^{(N+1)}$
(Eq.(\ref{invrelations4})).

The factorization for the tensor $\Phi^{(N+1)}$ is then simply given
by the $(N+1)^\text{th}$ equation in Eqs.[15]. The expression for
the $\Phi^{(N+1)}$ tensor contains only kinetic tensors of order $n
\le N$ , as well as the $(N+1)^\text{th}$ order cumulant. Once this
cumulant is known the system of Eqs.(8) closes and one arrives at
the $N^\text{th}$ order tensor DFT theory.

\subsection{$N=3$ Hydrodynamic tensor-DFT} At the $N=3$ level the
factorization of $\Phi^{(4)}$ is given by Eq.(\ref{invrelations4})
with $\phi^{(4)}$ set to zero, or:
\begin{eqnarray}
\Phi^{(4)}&=&4\rho^{-1}\overline{J \Phi^{(3)}} -
12\rho^{-2}\overline{J J
\Phi^{(2)}}\nonumber\\&&+6\rho^{-3}\overline{J J J J}\nonumber\\&& +
3\rho^{-1}\overline{\Phi^{(2)} \Phi^{(2)}} -3\rho \overline{e^{(2)}
e^{(2)}} +\rho e^{(4)} + \rho \phi^{(4)} \label{Phiphi4}.
\end{eqnarray}
In order to complete the theory we need to obtain $\phi^{(4)}$. For
this, we study the static linear response of a homogeneous electron
gas. In the ground state all the odd order $\Phi^{(\alpha)}$-tensors
vanish (a ground state has no currents as its wave-function is real
when there is no magnetic field and no degeneracy), so that the
first three terms in Eq.(\ref{Phiphi4}) would give only non-linear
contributions and can be neglected. As a result, the required
factorization for static studies simplifies as (here we restore the
indices):
\begin{eqnarray}
\Phi^{(4)}_{ijkl}&=& 3\rho^{-1}\overline{\Phi^{(2)}_{ij}
\Phi^{(2)}_{kl}}\nonumber\\&&+3
\rho(c_4(\rho)-c_2^2(\rho))\overline{\delta_{ij} \delta_{kl}}
+\phi^{(4)}_{ijkl}.
\end{eqnarray}

\section{Static linear response of homogeneous fermions and
adjustment to the Lindhard structure factor} \label{StaticResponse}
\begin{figure}[tb]
\includegraphics[scale=0.7]{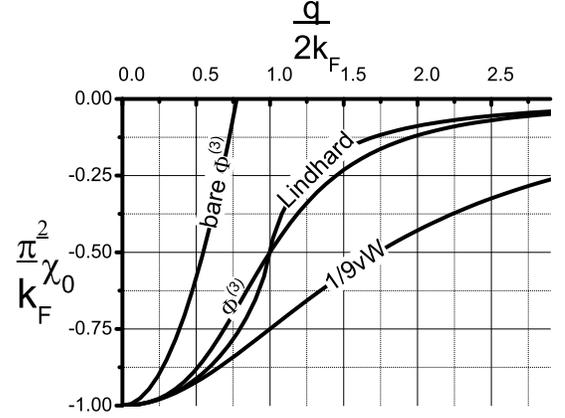}
\caption{\label{Figure1} The bare structure factors, $\chi_0$, on a
scaled momentum scale, $q/2k_F$, for the $1/9$-von Weiszacker
approach, bare and adjusted $\Phi^{(3)}$ HTDFT theories, and the
Lindhard function (free-electron gas static density-density
correlator). The $\Phi^{(3)}$ HTDFT is fitted to have the three
properties of the Lindhard function given in
Eq.(\ref{threeconditions}).}
\end{figure}
The static properties of a homogeneous electron liquid are
determined by the structure factor, $\chi(\bm q)$. The structure
factor is actually the static limit of the density-density
correlation function $\chi(\bm q)=\left.\langle\hat\rho(-\bm q,
-\omega)\hat\rho(\bm q, \omega)\rangle\right|_{\omega\to i0}$. The
physical meaning of $\chi(\bm q)$ is the ratio between the amplitude
of the infinitesimal harmonic change in electron density,
$\underline{\rho}(\bm q)$, and that of the external potential,
$\underline{v}_{ext}(\bm q)$, which induces the change in the
electron density:
\begin{subequations}
\begin{eqnarray}
\delta v_{ext}(\bm R) &=& \underline{v}_{ext} \exp(i \bm q \cdot \bm R) + c.c.,\\
\delta\rho(\bm R) &=& \underline{\rho}  \exp(i \bm q \cdot \bm R) + c.c.,\\
\underline{\rho}&=& \frac{\chi(\bm q)}{\rho_0} \underline{v}_{ext} +
C \underline{v}_{ext}^3 + \dots
\end{eqnarray}
\end{subequations}

In the ground state of a homogeneous liquid the non-zero values at
the $N=3$ level are the density $\rho_0$ and the second and the
fourth order dynamic tensors,
$\Phi^{(2)}_{ij}=\rho_0c_2(\rho_0)\delta_{ij}$,
$\Phi^{(4)}_{ijkl}=\rho_0c_4(\rho_0)\overline{\delta\cdot\delta}_{ijkl}$.
In a static linear response problem all the odd-order kinetic
tensors remain zero. Therefore, to study the static linear response
of the system we let the values of $\rho$, $\Phi^{(2)}$ and
$\Phi^{(4)}$ vary harmonically in space around their stationary
values:
\begin{subequations}
\begin{eqnarray}
v_\text{ext}&=&\underline v_\text{ext}
e^{i \bm q\cdot \bm R}+c.c. ,\\
\rho&=&\rho_0+\left(\underline{\rho}
e^{i \bm q\cdot \bm R}+c.c.\right),\\
\Phi^{(2)}_{ij} &=& c_2 \rho_0\delta_{ij}+\left(\underline{\Phi}^{(2)}_{ij}e^{i \bm q\cdot \bm R}
+c.c.\right),\\
\Phi^{(4)}_{ijkl} &=& 3c_4 \rho_0
\overline{\delta_{ij}\delta_{kl}}+\left(\underline{\Phi}^{(4)}_{ijkl}
e^{i \bm q\cdot \bm R} +c.c.\right) \nonumber\\&&+ \rho_0\phi^{(4)}_{ijkl},\\
\nabla_i\tilde V &=& q_i \left(i (\tilde v(q)
\underline{\rho}+\underline{v}_{ext})e^{i \bm q\cdot \bm
R}+c.c.\right),
\end{eqnarray}
\end{subequations}
where the underlined variables are the linear response coefficients, while
\begin{eqnarray}
\tilde v(q) = \frac{4\pi}{q^2} - \frac{\partial
V_{xc}(\rho)}{\partial \rho}(\rho_0).
\end{eqnarray}
With the use of Eq.(\ref{Phiphi4}) the infinitesimal deviation of $\Phi^{(4)}$ has the following form:
\begin{eqnarray}
&&\underline{\Phi}^{(4)}_{ijkl} = 3 D \underline{\rho}
\overline{\delta_{ij}\delta_{kl}} + 6 c_2
\overline{\delta_{ij}\underline{\Phi}^{(2)}_{kl}}+\rho_0\phi^{(4)}_{ijkl},
\end{eqnarray}
where
\begin{eqnarray*}
D = \left.\frac{\partial(\rho (c_4 - c_2^2))}{\partial
\rho}\right|_{\rho=\rho_0} - c_2^2 = -\frac1{15}k_F^4.
\end{eqnarray*}
Next we consider what terms can be in $\phi^{(4)}$.  Our purpose is
to make sure that the static response in the non-interacting case
would resemble the Lindhard function (static density response of
free electrons). The terms added should include the derivative of
the available quantities, \emph{i.e.}, the density and the stress
tensor, so that they will be vanishing for uniform densities.
Further, since $\phi^{(4)}$ is a fourth-order tensor, it needs to be
constructed from available tensors; the only ones available in the
static limit are $\nabla_i$, $\rho$, $\Phi_{kl}$ and $\delta_{jl}$.
It is easy to see by inspection that only the following local terms
are available to first order in the perturbation and to lowest
orders needed in $\nabla_i$:
\begin{widetext}
\begin{eqnarray}
\label{additionalterms} \rho_0\phi^{(4)}_{ijkl}(\bm R) = \Lambda
\overline{\nabla_i\nabla_j\nabla_k\nabla_l} \rho - 6 f
\overline{\nabla_i\nabla_j \Phi^{(2)}_{kl}} - 6 c_2 h
\overline{\delta_{ij} \nabla_k\nabla_l}\rho,
\end{eqnarray}
where $\Lambda, f$ and $h$ are dimensionless parameters. Even in
linear response these terms can be augmented by terms involving
further derivatives, e.g., terms involving a Laplacian of the
components in Eq. (\ref{additionalterms}), (i.e., $q^2$ in Fourier
space) but as orbital-free methods should be primarily geared
towards the long-wavelength limit, we do not consider here such
higher order terms in $q$.

The additional terms yield the following relation between the linear
response coefficients $\underline{\rho}$, $\underline{\Phi}^{(2)}$,
and $\underline{\Phi}^{(4)}$:
\begin{eqnarray}
\underline{\Phi}^{(4)}_{ijkl}(\bm q) = 6 \overline{\left( c_2
\delta_{ij}+f q_i q_j\right)\underline\Phi^{(2)}_{kl}} + 3 D
\underline\rho\overline{\delta_{ij}\delta_{kl}} + \Lambda
\underline\rho q_i q_j q_k q_l + 6 c_2 h \underline\rho
\overline{q_iq_j\delta_{kl}}. \label{Phi4linearizedvariation}
\end{eqnarray}
Finally, the linearized equations read:
\begin{subequations}
\begin{eqnarray}
&&q_\alpha \underline\Phi^{(2)}_{i\alpha} + q_i \rho_0\tilde v \underline\rho
+ q_i\rho_0 \underline v_{\text{ext}}=0,
\end{eqnarray}
and
\begin{eqnarray}
&&3\overline{(c_2\delta_{ij}+fq_iq_j)(q_\alpha\underline
\Phi^{(2)}_{k\alpha})} + 3(c_2 + f q^2)\overline{q_i
\underline\Phi^{(2)}_{kl}}\nonumber\\
&& + \left( 3\left(D + c_2 \rho_0\tilde v + h c_2
q^2\right)\overline{\delta_{ij} q_k} +\left(\frac14\rho_0\tilde
v + \Lambda q^2 + 3 h c_2\right)q_iq_jq_k \right) \underline\rho\nonumber \\
&&+\left(3 c_2 \overline{\delta_{ij} q_k} +
\frac14q_iq_jq_k\right)\rho_0\underline v_{\text{ext}}=0.
\end{eqnarray}
\end{subequations}
where the index $\alpha$ is summed over. The only preferential
direction in the problem is the momentum vector, $\bm q$, so that
the dynamic tensor, $\underline\Phi^{(2)}$, can be decomposed into
the two form-factors:
\begin{eqnarray*}
\underline\Phi^{(2)}_{ij} = \delta_{ij}\underline \Phi^{(0)} +
\frac{q_iq_j}{q^2}\underline \Phi^{(1)}.
\end{eqnarray*}
Upon substituting this resolution into the initial equations and
equating independent spatial tensor components we arrive at three
equations for $\underline\rho$, $\underline\Phi^{(0)}$ and
$\underline\Phi^{(1)}$:
\begin{eqnarray*}
&&
\underline\Phi^{(0)}+\underline\Phi^{(1)}+\rho_0\tilde v \underline
\rho = - \rho_0 \underline v_{\text{ext}},\\
&&\left(-\rho_0 \underline v_{\text{ext}} - \rho_0\tilde v
\underline \rho\right) + \left(1 + \frac
f{c_2}q^2\right)\underline\Phi^{(0)} +
\left(-\frac{k_F^2}{3}+\rho_0\tilde v + hq^2\right)\underline \rho =
-\rho_0\underline v_{\text{ext}}, \\ &&3f\left(-\rho_0\underline
v_{\text{ext}} - \rho_0\tilde v \underline\rho\right) +
3\frac{c_2}{q^2}\left(1 + \frac f{c_2}q^2\right)\underline\Phi^{(0)}
+ \left( \frac14\rho_0\tilde v + \Lambda q^2 +
3hc_2\right)\underline \rho = -\frac14\rho_0\underline
v_{\text{ext}}.
\end{eqnarray*}
\end{widetext}
Upon solving these linear equations one gets:
\begin{subequations}
\begin{eqnarray}
\underline\rho &=& -\chi \rho_0 \underline v_{\text{ext}},\\
-\chi^{-1} &=& -\chi_0^{-1} + \tilde v, \label{renormalization}
\end{eqnarray}
where
\begin{eqnarray}
-\frac{\pi^2}{k_F}\chi_0 &=& \frac{1+20(2f-1/12)\eta^2}{1 -24 h
\eta^2 - 80\Lambda \eta^4},
\end{eqnarray}
\end{subequations}
and we introduced the dimensionless momentum:
\begin{eqnarray*}
\eta = \frac q{2k_F}.
\end{eqnarray*}
Eq.(\ref{renormalization}) is the definition of the structure factor
renormalized with respect to two-body interactions. Therefore
$\chi_0$ should be the structure factor of non-interacting
electrons. In order to adjust our theory to the realistic
description of electrons one should compare $\chi_0$ to the Lindhard
function:
\begin{eqnarray}
-\frac{\pi^2}{k_F}\chi_{\text{Lind}} = \frac12 +
\frac{1-\eta^2}{4\eta}\ln\left|\frac{1+\eta}{1-\eta}\right|.
\end{eqnarray}
The freedom in choosing the parameters $\Lambda, f$, and $h$ allows
us to fit our structure factor to the Lindhard function. The
Lindhard function has the following properties:
\begin{eqnarray}
-\frac{\pi^2}{k_F}\chi_{\text{Lind}} = \left\{
  \begin{matrix}
    \frac13\eta^{-2}, & \eta\to \infty \\
    1-\frac13\eta^2, & \eta\to 0 \\
    \frac12, & \eta=1 \\
  \end{matrix}. \label{threeconditions}
\right.
\end{eqnarray}
In order for our function, $\chi_0$, to possess these properties we
should choose:
\begin{eqnarray}
\Lambda = -\frac1{80}, f = \frac1{20}, \text{ and } h = -
\frac1{36}. \label{coefficients}
\end{eqnarray}
A comparison between the resulting structure factor of the proposed
theory with the Lindhard function and the structure factor provided
by the 1/9-von Weiszacker theory is given in Fig.1.

\section{Application to the ground state problem}
\label{nonhomogeneousstudy} We applied the $\Phi^{(3)}$-theory to a
ground state study of a non-perturbative non-homogenous jellium. We
chose a spherically symmetric infinite electron system in the
following positive jellium background density profile:
\begin{eqnarray}
\rho_0(r) &=& \rho_\infty + \Delta\rho(r),\label{density}\\
\Delta\rho(r) &=&
-\zeta\rho_\infty\left(1-\frac{r^2}{3r_0^2}\right)e^{-\frac{r^2}{2r_0^2}},\nonumber
\end{eqnarray}
where we took $\rho_\infty=0.01$, $\zeta=0.9$, and $r_0=3\text{ and
}2$ (all in a.u.). The additional non-homogeneous part of jellium
density $\Delta\rho$ integrates to zero so we avoid complications
connected with an overall non-neutral system. Alternatively this
system can be viewed as having constant jellium background density,
$\rho_\infty$, but with an external Gaussian potential:
\begin{eqnarray}
V_b(r) = r_0^2\zeta\rho_\infty\frac{4\pi}3e^{-\frac{r^2}{2r_0^2}},
\end{eqnarray}
which is related to $\Delta\rho(r)$ by the Poisson equation:
\begin{eqnarray}
\frac1{r^2}\frac\partial{\partial r}r^2\frac\partial{\partial r} V_b(r) = 4\pi\Delta\rho(r).
\end{eqnarray}
The Dirac exchange is used here:
\begin{eqnarray}
V_{xc}(\bm R) = \frac{(3\pi^2)^{1/3}}{\pi}\rho(\bm R)^{1/3}.
\end{eqnarray}
No correlation energy was employed (its contribution is very small;
it will be included in future studies).
\begin{figure*}[tb]
\includegraphics[scale=0.7]{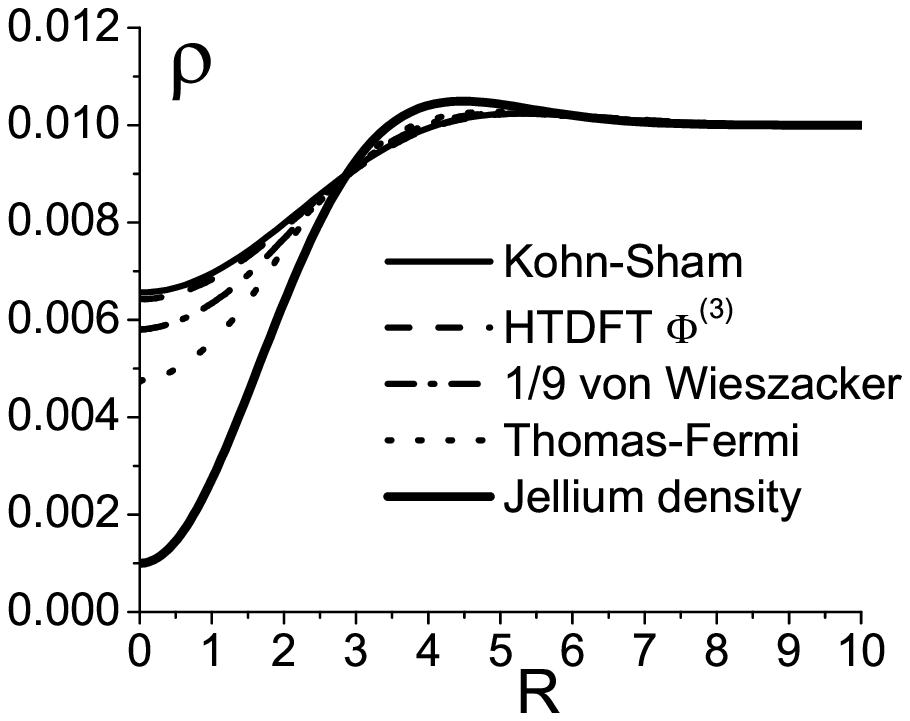}
\includegraphics[scale=0.7]{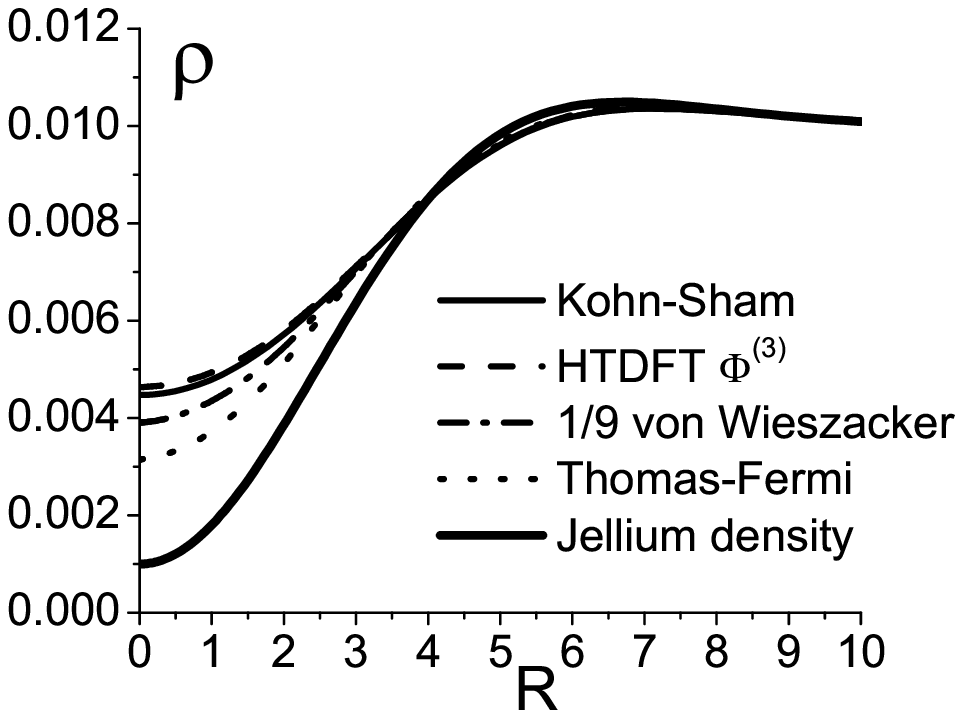}
\caption{\label{Figure2} The electron density profiles for a jellium
model with background positive density (lower solid line) given by
Eq.(\ref{density}) with $\zeta=0.9$ and $r_0 = 2, 3$ (left and right
graphs respectively) for Thomas-Fermi, $1/9$-von Weiszacker,
$\Phi^{(3)}$ HTDFT, and the Kohn-Sham orbital based approaches. All
quantities are in a.u.}
\end{figure*}
The simulations were performed by adiabatical turning on the
nonhomogeneous part of the jellium positive background density,
$\Delta\rho$. Initially the electron and the jellium densities are
homogeneous, $\rho_\infty$. The odd-order kinetic tensors, $J_i$ and
$\Phi^{(3)}_{ijk}$, are zero and the even order tensors are those of
a homogeneous electron liquid $\Phi^{(2)}_{ij}=\rho_\infty
e^{(2)}_{ij}$ and $\Phi^{(4)}_{ijkl}=\rho_\infty e^{(4)}_{ijkl}$
with the $e$-tensors given in Eq.(\ref{etensors}).

We then propagate the set of Eqs.(\ref{setofequations}) while the
jellium density gradually changes from homogenous to the final
$\rho_0(r)$; this ensures that the system remains at the
ground-state for all times. We implemented the adiabatic density by
setting
\begin{eqnarray*}
\rho_0(R,t) = (1-g(t))\rho_0(R) + g(t)\rho_\infty,
\end{eqnarray*}
where $g(t)$ is a smooth function rising from 0 to 1; we chose here,
quite arbitrarily,
\begin{eqnarray*}
g(t) = \frac1{1+\exp\left(\left(\frac{t_0-t}{\tau}\right)^3\right)},
\end{eqnarray*}
and used
\begin{eqnarray*}
t_0=3\tau.
\end{eqnarray*}
The width parameter, $\tau$, was typically taken as $50 a.u.$; this
value was more than sufficient for adiabatic convergence.
$\rho_0(R,t)$ is then used for the definition of the time-dependent
potential, Eq.(\ref{EffectivePotential}).

The evolution of the system is then determined from the four first
equations in (\ref{setofequations}), with the $\Phi^{(4)}$ tensor
given by Eq.(\ref{Phiphi4}) and Eq.(\ref{additionalterms}). The 3D
equations were discretized and the derivative were evaluated by
Fourier-transforms, as was the Coulomb integral. Grid spacings of
1.6 a.u. - 2 a.u. were sufficient to converge when the hole width
parameter, $r_0$, was set at 2.0 or 3.0 a.u., respectively. A simple
fixed step Runge-Kutta algorithm with dt=0.2 a.u. was used to evolve
the equations in time.

We compared the results to Thomas-Fermi, von-Weiszacker, and
plane-wave Kohn-Sham simulations. The latter were done by a standard
plane wave code; interestingly, we found that the grid spacing
needed to converge the Kohn-Sham plane wave simulations had to be
smaller by about 20$\%$ than those needed in the HFDFT code, so that
they were about 1.3 and 1.6 a.u. for $r_0$=2.0 and 3.0,
respectively.  The grids contained typically $(20)^3$ points.

Fig.2 shows that HTDFT gives essentially the same density as the
Kohn-Sham approach, while the von-Weiszacker and Thomas-Fermi
results deviate significantly.  Since the two-body interaction is
treated the same in all four simulations, this proves that the
hydrodynamic approach yields, even for this system which is shifted
strongly away from uniformity, the same densities as the essentially
exact description of the kinetic energy in the Kohn-Sham approach.

\section{Time-dependent linear response and the collective modes}
\label{dynamicresponse} In our previous paper \cite{ours} we studied
the ground-state of a homogenous electron gas at the $N=2$ level,
with the assumption that $\phi^{(N+1)}$ is zero.  Here we extend the
studies to $N=3$, with $\phi^{(4)}$, as given by
Eqs.(\ref{additionalterms}),(\ref{coefficients}).  We derive the
governing formulae in general, and arrive at analytical limits in
the long wavelength limit (where $\phi^4$ is not contributing),
showing new kinds of excitations.

In the ground state of a homogeneous liquid the non-zero values at
the $N=3$ level are the density $\rho_0$ and the second and the
fourth order dynamic tensors,
$\Phi^{(2)}_{ij}=\rho_0c_2(\rho_0)\delta_{ij}$,
$\Phi^{(4)}_{ijkl}=\rho_0c_4(\rho_0)
\overline{\delta_{ij}\delta_{kl}}$. To study the linear response of
the system we let all the values in the problem vary harmonically
around their stationary values:
\begin{subequations}
\begin{eqnarray}
\rho&=&\rho_0+\left(\underline\rho
e^{-i(\omega t - \bm q\cdot \bm R)}+c.c.\right),\\
J_i &=& \underline{J}_i
e^{-i(\omega t - \bm q\cdot \bm R)}+c.c.,\\
\Phi^{(2)}_{ij} &=& c_2
\rho_0\delta_{ij}+\left(\underline\Phi^{(2)}_{ij}
e^{-i(\omega t - \bm q\cdot \bm R)}+c.c.\right),\\
\Phi^{(3)}_{ijk} &=& \underline\Phi^{(3)}_{ijk}
e^{-i(\omega t - \bm q\cdot \bm R)}+c.c.,\\
\nabla_i\tilde V &=& q_i \tilde v(q) \left(i \underline\rho
e^{-i(\omega t - \bm q\cdot \bm R)}+c.c.\right),
\end{eqnarray}
\end{subequations}
\begin{widetext}
After linearizing Eqs.(\ref{setofequations}) one gets:
\begin{subequations}
\label{linearizedeqs}
\begin{eqnarray}
\omega\underline{\rho} &=&q_\alpha\underline J_\alpha,\\
\omega \underline J_{i}&=&q_\alpha\underline\Phi^{(2)}_{i\alpha}+\rho_0 q_i \tilde v(q) \underline\rho,\\
\omega\underline\Phi^{(2)}_{ij}&=&q_\alpha\underline\Phi^{(3)}_{ij\alpha},\\
\omega\underline\Phi^{(3)}_{ijk}&=&
3\overline{(c_2\delta_{ij}+fq_iq_j)(q_\alpha\underline
\Phi^{(2)}_{k\alpha})} + 3(c_2 + f q^2)\overline{q_i
\underline\Phi^{(2)}_{jk}}\nonumber\\&& + \left( 3\left(D + c_2
\rho_0\tilde v + h c_2 q^2\right)\overline{\delta_{ij} q_k}
+\left(\frac14\rho_0\tilde v + \Lambda q^2 + 3 h c_2\right)q_iq_jq_k
\right) \underline\rho,
\end{eqnarray}
\end{subequations}
where the linearized variation of $\Phi^{(4)}$ is taken from
Eq.(\ref{Phi4linearizedvariation}), and $\alpha$ is again summed
over. This is a system of linear homogeneous equations, and to find
its solutions we have to diagonalize it.

All the variables in Eqs.(\ref{linearizedeqs}) could be expressed in terms of $\underline\rho$ and $\underline\Phi^{(2)}_{ij}$.
Therefore, we can consider the equations on $\underline\Phi^{(2)}$ and $\underline{\rho}$ only without losing any solutions. In
matrix form these equations read:
\begin{subequations}
\label{linearequations}
\begin{eqnarray}
\frac{\omega^2}{q^2} \underline{\rho} &=& \text{Tr}\left(\underline{\bm\Phi}^{(2)} \bm Q \right)+ \rho_0 \tilde v(q) \underline{\rho},\\
\frac{\omega^2}{q^2}\underline{\bm\Phi}^{(2)}&=&(c_2+fq^2)\left(\underline{\bm\Phi}^{(2)}+ 2\left\{\underline{\bm \Phi}^{(2)},\bm
Q\right\}\right) + (c_2\bm I+fq^2\bm Q) \text{Tr}\left(\underline{\bm\Phi}^{(2)} \bm Q \right) \nonumber
\\&& +\left(\bm I Z(q) + \bm Q Z'(q) \right)\underline{\rho},
\end{eqnarray}
\end{subequations}
\end{widetext}
where
\begin{eqnarray*}
Z(q) & = & D + c_2\rho_0\tilde v + h c_2 q^2,\\
Z'(q)& = &2 Z + \frac14\rho_0\tilde v(q) q^2 + \Lambda q^4 +
3hc_2q^2,
\end{eqnarray*}
and $\text{Tr}$ means matrix trace, $I_{ij}\equiv\delta_{ij}$ is the
$3\times 3$ unity matrix, $Q_{ij}=q_iq_j/q^2$, curly brackets denote
an anticommutator, and capital bold face letters refer here to
matrices. Without loss of generality we can always assume that the
wave-vector $\bm q$ is directed along the x-axis ($\bm q
=(q,0,0)^T$), so that
\begin{eqnarray}
Q=  \begin{pmatrix}
    1 & 0 & 0\\
    0 & 0 & 0\\
    0 & 0 & 0
  \end{pmatrix}.
\end{eqnarray}
There are several solutions for these equations. The first three solutions are decoupled from the density fluctuations so that
$\underline \rho = 0$ for all of them. They are:
\begin{subequations}
\begin{eqnarray}
\underline{\Phi}^{(2)}=  \begin{pmatrix}
    0 & 1 & 0\\
    1 & 0 & 0\\
    0 & 0 & 0
  \end{pmatrix},
\begin{pmatrix}
    0 & 0 & 1\\
    0 & 0 & 0\\
    1 & 0 & 0
  \end{pmatrix},\label{TransverseSound}
\end{eqnarray}
which corresponds to the dispersion relation
\begin{eqnarray}
\omega^2 = 3/5 k_F^2 q^2, \label{disrelation1}
\end{eqnarray}
and
\end{subequations}
\begin{subequations}
\label{thirdsolution}
\begin{eqnarray}
\underline{\Phi}^{(2)}=  \begin{pmatrix}
    0 & 0 & 0\\
    0 & 0 & 1\\
    0 & 1 & 0
  \end{pmatrix},
\end{eqnarray}
with dispersion
\begin{eqnarray}
\omega^2=1/5k_F^2q^2.\label{disrelation2}
\end{eqnarray}
\end{subequations}

The first two solutions, Eq.(\ref{TransverseSound}), correspond to transverse sound as the current is given as:
\begin{eqnarray}
\omega(q)\underline{J_i}=q_j\underline\Phi^{(2)}_{ij} =\begin{pmatrix} 0 \\ q \\
0 \end{pmatrix}, \begin{pmatrix} 0 \\ 0 \\ q \end{pmatrix}.
\end{eqnarray}
Note, however, that the velocity of this transverse sound mode is
different from the one found for the same mode within the $N=2$
theory \cite{ours}. The third solution in Eq.(\ref{thirdsolution})
is a new sound mode.  This mode involves neither density nor current
fluctuations and corresponds to transverse quadrupole fluctuations
of the Fermi sea.
\begin{figure}[tb]
\includegraphics[scale=0.7]{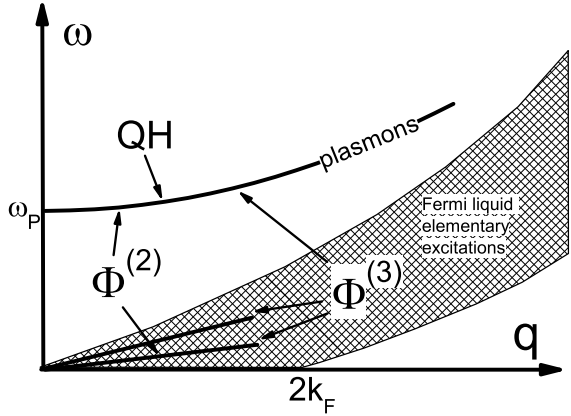}
\caption{\label{Figure3} The elementary excitation spectrum provided
by Quantum Hydrodynamics (QH), HTDFT $\Phi^{(2)}$ and HTDFT
$\Phi^{(3)}$ theories. QH gives only a plasmon mode. HTDFT also
gives transverse sound modes which mimic the RPA elementary
excitations in Fermi liquid. $\Phi^{(3)}$ HTDFT gives additional
sound modes with respect to $\Phi^{(2)}$ HTDFT confirming the
conjecture made in Ref.\cite{ours} that when increasing the order of
HTDFT new sound modes should appear, and they will gradually cover
the entire continuous RPA density of states of Fermi liquid.}
\end{figure}

The next three solutions are found by representing the tensor
$\Phi^{(2)}$ in terms of the remaining diagonal tensors ($I$ and
$Q$):
\begin{eqnarray}
\underline{\bm \Phi}^{(2)} = \alpha \bm Q +\beta (\bm I - \bm Q),
\end{eqnarray}
which leads, upon insertion into Eqs.(\ref{linearequations}), to the
following equations for $\alpha,\beta,\underline{\rho}$:
\begin{subequations}
\begin{eqnarray}
\frac{\omega^2}{q^2} \alpha &=& 6 \left( c_2 + f q^2 \right)\alpha + \left(Z(q) + Z'(q))\right)\underline\rho .\\
\frac{\omega^2}{q^2} \beta  &=& c_2 \alpha +\left( c_2 + f q^2 \right)\beta +  Z(q)\underline\rho , \\
\frac{\omega^2}{q^2}\underline{\rho} &=& \alpha + \rho_0 \tilde v(q)
\underline\rho .
\end{eqnarray}
\end{subequations}
This set of equations has complicated solutions, which, however,
could be simplified in low-wavelength limit. In this limit, we can
leave only the leading terms in $q$; in the effective potential it
is the divergent Fourier components of the Coulomb potential. In the
long-wavelength limit the system of equations has the following
form:
\begin{eqnarray}
\begin{pmatrix}
    6 & 0 & 3\\
    1 & 1 & 1\\
    a(q) & 0 & a(q)
  \end{pmatrix}
\begin{pmatrix}
    \alpha\\
    \beta\\
    \underline{\rho}'
  \end{pmatrix}=
 \omega'^2
  \begin{pmatrix}
    \alpha\\
    \beta\\
    \underline{\rho}'
  \end{pmatrix},
\end{eqnarray}
where $a(q)=4\pi \rho_0 /(q^2 c_2)$, $\omega'^2=\omega^2/(q^2c_2)$,
$\underline{\rho}'=4\pi\rho_0\underline{\rho}/q^2$, and $a(q)\gg1$.
Dropping the terms of order $a(q)^{-1}$ and smaller, the three
eigenvalues and corresponding eigenvectors are:
\begin{subequations}
\begin{eqnarray}
\omega^2&=&\frac15k_F^2 q^2,\quad (\alpha, \beta, \underline{\rho}') = ( 1, 2/3 , -1);\label{fourthmode}\\
\omega^2&=&\frac35k_F^2 q^2,\quad (\alpha, \beta, \underline{\rho}') = ( 1,  0  , -1);\label{fifthmode}\\
\omega^2&=&\omega_P^2+\frac35k_F^2 q^2,\quad (\alpha, \beta,
\underline{\rho}') = (0, 0, 1),
\end{eqnarray}
\end{subequations}
where $\omega_P^2=4\pi\rho_0$ is the plasmon frequency. Note that
the first two of the three modes (\ref{fourthmode}, \ref{fifthmode})
have the same eigenvalues as the transverse modes in
Eqs.(\ref{disrelation1}) and (\ref{disrelation2}).

The total spectrum given by $N=3$ HTDFT for elementary excitations
in the homogeneous electron gas is given in Fig.(\ref{Figure3}). The
spectrum found differs from that of the $N=2$ approach by an
additional sound mode with velocity $\sqrt{3/5} k_F$ and by shifting
the previous sound modes from $\sqrt{3/5}k_F$ to $\sqrt{1/5}k_F$.
This result confirms the conjecture made in Ref.\cite{ours}, that
with increasing $N$ new sound modes should appear, and that they
will gradually cover the entire continuous RPA density of states in
the Fermi liquid.

\section{Conclusions}
In conclusions, we have shown that HTDFT can also be used for
time-independent studies. We have supplanted our previous conjecture
where we assumed that the terms in the equation of motion hierarchy
should be terminated with the next relevant cumulant (\emph{i.e.},
$\phi^{(N+1)}$) being zero; instead, we now derived $\phi^{(N+1)}$
from fitting the linear response to a HEG. The resulting set of
equations (given at the $N=3$ level by Eqs.
(\ref{additionalterms}),(\ref{additionalterms}),(\ref{setofequations}),
and (\ref{Phiphi4}) ) is closed and can be propagated forward in
time.

The linear response in the static limit is fitted to the Lindhard
function HEG for both short, intermediate and long wavelengths (for
comparison, the 1/9 in the von-Weiszacker approach is obtained to
fit long wavelengths, while a fit to long wavelengths would have
required replacing the 1/9 by 1 in the von-Weiszacker theory).  We
have then applied HTDFT away from equilibrium, for a case of a
jellium density with a deep hole in the middle, and have shown
excellent agreement with the Kohn-Sham results, in a case where more
approximate theories such as Thomas Fermi and von-Weiszacker fail;
this is directly due to the fact that their structure factor do not
follow the Lindhard function except at low wavelengths.

The last part of the paper dealt with time-dependent linear response
studies at the present level, $N=3$. The analytical studies have
confirmed our previous assertion, that as the level of the theory
increases, more and more transverse excitations are found. A new
excitation at the N=3 level couples neither to the current nor to
the density.  All excitations, including the new ones, lie within
the RPA density of stats of elementary excitations in a Fermi
liquid.

Future work will study the applicability of the approach to covalent
chemical systems, where the directionality of the tensors should
enable a correct description even at a low N, possibly as low as
$N=3$.  Further, dynamic susceptibilities will be studied so that
further terms, depending on $J$, $\Phi^{(3)}$, \emph{etc.}, will be
included in the terminating cumulant ($\phi^4$ here, $\phi^{(N+1)}$
in general) so that the theory will be valid over a wide range of
frequencies and wavevectors. In addition, effects of magnetic fields
are straightforward to incorporate. The basic formalism developed
here and in forthcoming work will then enable the application to
dynamical problems which straddle the transition between molecular
and nanostructures.

We note that other applications to fermionic systems can also be
envisioned.  For example, by replacing the zeroth-order HEG density
matrix with a temperature-dependent density matrix, and fitting the
coefficients of the derivative terms in the cumulant to a
temperature dependent Lindblad expression, we will get a
temperature-dependent HTDFT theory which can be applied to plasmas
and to studies of narrow conduction bands. Similarly, applications
to nuclear systems can also be envisioned.

Other future improvements will include better methods to solve the
time-dependent HTDFT equations. One approach will be to include
external electric fields that will have dipole and quadruple (or
higher) components that will be time-dependent.  The electric fields
will be chosen, at each time-instant, to remove energy from the
system (\emph{i.e.}, to reduce the trace of $\Phi^{(2)}$ plus the
total potential).

\section{Acknowledgements}
We thank Roi Baer for his help in performing and analyzing the
numerical and linear response studies. This work was supported by
the NSF and PRF. We are thankful for Emily Carter and the Referee
for helpful comments.

\newpage \printfigures
\end{document}